\documentclass[sigconf]{acmart}

\AtBeginDocument{%
  }

\usepackage{graphicx}
\usepackage{subcaption}
\usepackage{xcolor}

\newif\ifshowreviewchanges
\showreviewchangestrue

\copyrightyear{2026}
\acmYear{2026}
\setcopyright{cc}
\setcctype{by}
\acmConference[MODELS Companion 2026]{ACM/IEEE 29th International Conference on Model Driven Engineering Languages and Systems}{October 04--09, 2026}{M\'alaga, Spain}
\acmBooktitle{ACM/IEEE 29th International Conference on Model Driven Engineering Languages and Systems (MODELS Companion 2026), October 04--09, 2026, M\'alaga, Spain}
\acmDOI{10.1145/3837062.3839341}
\acmISBN{979-8-4007-2903-4/2026/10}
\begin{document}

\title{Quantifying the Relationship Between Clinical Safety and Environmental Impact in Therapeutic LLMs}

\author{Alireza A. Safaei}
\affiliation{%
  \institution{University of Isfahan}
  \city{Isfahan}
  \country{Iran}
}
\email{alirezaakhavansafaei@eng.ui.ac.ir}

\author{Laura M. Vowels}
\affiliation{%
  \institution{University of Roehampton}
  \city{London}
  \country{UK}}
\email{laura.vowels@roehampton.ac.uk}

\author{Matthew J. Vowels}
\affiliation{%
 \institution{Kivira Health}
 \city{New York City}
 \state{New York}
 \country{U.S.A.}}
 \email{matt@kivira.health}

 \author{Apoorv Jha}
\affiliation{%
 \institution{Kivira Health}
 \city{New York City}
 \state{New York}
 \country{U.S.A.}}
 \email{apoorv.jha@icloud.com}

\author{Shekoufeh Rahimi}
\affiliation{%
 \institution{University of Roehampton}
  \city{London}
  \country{UK}
  }
\email{shekoufeh.rahimi@roehampton.ac.uk}

\renewcommand{\shortauthors}{Safaei et al.}

\begin{abstract}
The deployment of large language models (LLMs) in mental health contexts raises questions about the relationship between clinical safety and environmental cost. In this paper, we examine this relationship by combining K-Bench clinical safety scores with EcoLogits life-cycle assessment estimates across 47 supported model configurations. We evaluate model performance and environmental impact across four dimensions: energy use, carbon emissions, water consumption, and abiotic depletion. The results indicate a non-linear trade-off at the upper end of the safety distribution: a 2.61 percentage-point increase in clinical safety score corresponded to an approximately 60-fold increase in estimated energy use per million output tokens. Row-level analyses further suggest that additional test-time compute did not consistently improve clinical safety and, in some configurations, was associated with lower clinical safety scores. These findings suggest that relying solely on larger models or additional inference-time computation may be an inefficient strategy for improving safety in therapeutic AI systems. We discuss the implications for sustainable deployment and highlight dynamic model selection, including model cascading, as a potential approach for reducing environmental impact while preserving clinical performance in higher-risk cases.
\end{abstract}




\maketitle

\section{Introduction}
Large language models (LLMs) are increasingly being explored for use in digital healthcare, including as conversational agents for mental health support \cite{maity2025llmhealthcare, yang2026medmt, jin2025llmmentalhealth, he2023conversationalagents}. These systems may offer a scalable way to provide context-sensitive support, but their use in clinical or quasi-clinical settings also raises important safety concerns. A therapeutic conversational agent must be able to sustain multi-turn dialogue, identify clinically relevant risk signals, and respond appropriately to domains such as suicidal ideation or self-harm, while avoiding responses that could cause harm or undermine care \cite{bentley2026veramh, medrxiv2026suicide, dwyer2025mindbenchai}. Understanding how model choice affects this balance is therefore an important problem for developers and healthcare organizations.

A common approach uses frontier-scale models, such as \texttt{gpt-5.5} or \texttt{claude-opus-4.8}, assuming larger and more capable models better handle clinical nuance \cite{wind2026scalinglawsclinicalllms, malgaroli2025large}. These models often perform strongly on clinical safety benchmarks, but they also entail higher inference costs. Serving large models at scale requires substantial compute, which has implications for operational energy use and associated environmental impacts, including global warming potential \cite{elsworth2025measuring}.

Clinical AI safety and environmentally sustainable AI inference are usually evaluated separately. As a result, model selection is often guided by safety or capability benchmarks without estimating their environmental cost. This makes it difficult to assess whether marginal improvements in clinical safety are associated with proportionate increases in energy use, carbon emissions, water consumption, or other life-cycle impacts. We quantify the relationship between clinical safety performance and environmental inference cost across LLM architectures.

To do so, we combine the public K-Bench clinical safety leaderboard with sustainability estimates from the EcoLogits package \cite{Rincé2025}. The initial dataset included 90 therapeutic model configurations, of which 47 had sufficient model support to estimate standardized environmental metrics. We analyze these configurations across clinical safety scores and four environmental indicators: energy use, global warming potential, water consumption, and abiotic depletion. This analysis identifies a Pareto frontier in which the highest safety scores are associated with disproportionately larger environmental footprints. The results provide evidence that smaller, optimized models may offer a more efficient deployment option for many therapeutic AI use cases, while larger models may be better reserved for cases where their additional safety performance justifies the added cost.

\section{Background and Related Work}
Recent work on therapeutic AI has increasingly used empirical, multi-turn evaluations to assess model safety \cite{bentley2026veramh}. Single-turn NLP benchmarks are often poorly suited to this setting because clinical conversations are sequential, context-dependent, and sensitive to changes in user state over time \cite{yang2026medmt, gong2026meddialogrubrics, liu2026jmedethicbench}. In longer interactions, models may lose relevant context, drift from earlier instructions, or respond inconsistently to clinically important information \cite{yang2026medmt, dutta2026mental}. Several studies therefore use simulated or adversarial dialogue to evaluate responses to higher-risk scenarios, including suicidal ideation and self-harm \cite{medrxiv2026suicide, arnaizrodriguez2026between}.

K-Bench uses factorial designs to examine how model performance changes as risk factors and comorbidities accumulate. Existing evaluations suggest that larger or more capable models often achieve higher safety scores, although the relationship between model scale, clinical performance, and deployment cost remains incompletely characterized \cite{malgaroli2025large, olisaeloka2026safety, stamatis2026beyond}. At the same time, safety systems must avoid excessive risk sensitivity. Overly cautious responses may interrupt otherwise appropriate conversations or escalate cases unnecessarily, creating a trade-off between risk detection and conversational continuity \cite{medrxiv2026suicide}. One proposed approach is architectural separation, in which dedicated risk-detection components monitor dialogue state alongside the main conversational model \cite{dutta2026mental}. However, these frameworks generally omit resource use and environmental impact.

Green AI work has developed methods to estimate machine-learning environmental footprints. Early studies focused primarily on the carbon cost of model training \cite{luccioni2023estimating}, but recent work emphasizes inference because deployed generative models may be queried repeatedly at scale \cite{elsworth2025measuring}. This has led to increased interest in estimating energy use and carbon emissions during inference across different model architectures and deployment settings \cite{luccioni2024power, oviedo2026energyuseaiinference}.

Inference costs are not uniform across the generation process. Hardware profiling indicates that autoregressive decoding can be more energy-intensive per token than initial prompt processing, in part because generation is constrained by memory bandwidth and sequential token production \cite{faiz2024llmcarbon, luccioni2024power}. This distinction is relevant for therapeutic applications, where conversations may be long and where some systems use additional reasoning steps, hidden tokens, or agentic workflows. Empirical estimates suggest that such approaches can substantially increase energy use per query compared with simpler inference patterns \cite{jegham2025howhungry, oviedo2026energyuseaiinference}. Frameworks such as EcoLogits \cite{Rincé2025} estimate these impacts across model endpoints using hardware, data center efficiency, and regional electricity mixes. These estimates can include energy use, global warming potential, and related life-cycle indicators \cite{fu2024llmco2, jegham2025howhungry, ozcan2025quantifying}.

A third relevant line of work concerns multi-objective optimization. Green AI argues for evaluating efficiency alongside task performance \cite{Schwartz2020GreenA}. This framing is especially relevant when marginal performance gains require disproportionately larger amounts of computation. Work on compute-optimal scaling has also shown that larger parameter counts are not always the most efficient route to improved model performance, and that smaller models can perform competitively when appropriately trained and optimized \cite{Hoffmann2022TrainingCL}.

Recent approaches apply similar reasoning to LLM deployment through model routing or dynamic model cascading. In these systems, lower-cost models handle routine queries, while larger models are reserved for cases that are expected to require additional capability \cite{Barros2025SmallIS, Jin2025EnergyCost}. Pareto analysis formalizes this trade-off by identifying the best observed performance for a given resource budget. For therapeutic AI, such methods may help distinguish cases where the added environmental cost of a larger model is justified from cases where a smaller model provides comparable performance.

Although clinical safety benchmarking, environmental impact estimation, and Pareto optimization have each developed as active research areas, they are rarely combined in evaluations of therapeutic AI. In particular, there is limited evidence on how clinical safety scores relate to estimated inference footprints across models. This study addresses that gap by combining K-Bench clinical safety data with EcoLogits sustainability estimates, allowing us to examine the safety–sustainability trade-off in therapeutic LLM deployment.

\section{Methodology}

\textbf{Dataset Curation and K-Bench:} To evaluate the relationship between clinical safety and environmental impact, we used the public K-Bench leaderboard. K-Bench (\url{https://k-bench.ai}) is a transcript-based benchmark for therapist-style AI systems. It evaluates model behaviour in simulated multi-turn mental health conversations, with particular emphasis on risk recognition, risk exploration, supportive conversation, ethical reasoning, and appropriate boundaries. The leaderboard contains 90 evaluated model--prompt--reasoning configurations across 28 base models and 13 provider namespaces.

K-Bench is built from a structured synthetic vignette pipeline. Vignettes are generated using a factorial design over four risk domains: suicide, self-harm, domestic violence, and substance misuse. The generator combines all domain combinations and a no-risk condition with three disclosure levels: low, moderate, and high. This produces 48 experimental cells. The 14{,}400-row vignette pool includes demographic, adverse childhood experience, presenting-problem, conversational-style, protective-factor, and domain-specific risk variables. Each vignette becomes a simulated patient prompt for a multi-turn conversation with the evaluated model.

The benchmark uses a fixed 200-vignette evaluation cohort for public leaderboard scoring. Each leaderboard row corresponds to a specific model configuration, defined by the base model, prompt variant, provider prompt version, and provider reasoning setting. Scores are reported on a 0--100 scale. The overall score summarizes performance across all rubric dimensions, while the risk score combines the two safety-critical dimensions: clinical judgement and risk awareness, and risk exploration. Because the present study focuses on clinical safety rather than general conversational quality, we used the combined risk score as the primary clinical outcome.

K-Bench scoring is based on a structured rubric originally calibrated against clinician ratings. Clinicians rated transcripts using domain-specific risk items and global response-quality items. The rubric contains risk-domain items for active safety domains and global items assessing ethical reasoning, supportive conversation, psychological knowledge, cultural and contextual competence, and autonomy and boundaries. Clinician consensus labels calibrated an automated judge. In the calibration set, the AI judge achieved Cohen's $\kappa = 0.850$ against clinician consensus, with 90.3\% raw item agreement and mean normalized dimension MAE of 0.040. The calibrated judge was then used to score the public evaluation cohort.

\textbf{Standardized Sustainability Estimation:} For the environmental analysis, we used the \texttt{ecologits} Python package (v0.11.0) to estimate the inference footprint of each supported K-Bench configuration. EcoLogits estimates environmental impact by combining model metadata with assumptions about the underlying hardware, data center efficiency, and electricity mix. Where available, model metadata are linked to hardware profiles such as NVIDIA H100 or A100 GPU deployments. Energy estimates are then adjusted using Power Usage Effectiveness (PUE) values and regional carbon-intensity assumptions for the relevant provider or default deployment setting, including the `electricity mix zone: provider default' option. We selected EcoLogits because commercial APIs do not expose the hardware and runtime data required by CodeCarbon or Green Algorithms. It enables consistent endpoint comparisons, but provides modeled estimates rather than direct measurements.

The resulting estimates include both operational and life-cycle components. Operational impacts cover inference energy, while life-cycle components include amortized hardware impacts. We extracted four environmental indicators: energy use (kWh), global warming potential (kgCO2eq), water consumption (L), and abiotic depletion potential (kgSbEq). To allow comparison across models and configurations, all indicators were normalized to 1,000,000 generated output tokens. This standardizes comparison across systems differing in response length, latency, or deployment assumptions.

\textbf{Data Synthesis and Filtering:} We merged the K-Bench clinical leaderboard with the EcoLogits sustainability estimates by matching each model to its corresponding endpoint configuration. Models were retained only when EcoLogits provided sufficient support to estimate environmental indicators using available model and provider metadata. This filtering step highlighted a limitation of current sustainability estimation for commercial and closed-weight systems. Of the 28 base architectures in the initial dataset, 15 could not be estimated because the required architectural, hardware, or routing information was unavailable. These unsupported models included \texttt{meta-llama/llama-4-maverick}, \texttt{deepseek/deepseek-v4-flash}, \texttt{anthropic/claude-fable-5}, and \texttt{x-ai/grok-4.20}. We excluded these models from the environmental analysis rather than estimating their impacts from incomplete information. The final dataset consisted of 47 configurations across 13 base model architectures.

\section{Results}

\textbf{Clinical Safety Scores and Environmental Impact:}  We first examined the relationship between clinical safety scores and estimated environmental impact. Table \ref{tab:models} shows a subset of base models, illustrating resource-use differences among relatively high-scoring models.

\begin{table*}[t]
\footnotesize
\caption{Clinical Safety vs. Estimated LCA Environmental Impact per 1M Output Tokens}
\label{tab:models}
\begin{tabular}{lrrrrr}
\toprule
\textbf{Base Model} & \textbf{Best Risk Score} & \textbf{Energy (kWh)} & \textbf{GWP (kgCO2eq)} & \textbf{Water (L)} & \textbf{ADPe (kgSbEq)} \\
\midrule
openai/gpt-5.5 & 96.02 & 6.4390 & 2.6041 & 23.2207 & 0.00000786 \\
openai/gpt-5.2 & 95.15 & 1.9819 & 0.8213 & 7.1474 & 0.00000353 \\
anthropic/claude-opus-4.8 & 94.34 & 4.5892 & 1.8539 & 17.2084 & 0.00000548 \\
anthropic/claude-haiku-4.5 & 93.41 & 0.1095 & 0.0455 & 0.4163 & 0.00000020 \\
google/gemini-3.5-flash & 90.43 & 4.8647 & 1.9099 & 19.6953 & 0.00000271 \\
mistralai/mistral-medium-3-5 & 87.68 & 0.5891 & 0.0343 & 3.5986 & 0.00000080 \\
openai/gpt-4o-mini & 85.27 & 0.0846 & 0.0375 & 0.3051 & 0.00000028 \\
\bottomrule
\end{tabular}
\end{table*}

The results show that the highest clinical safety scores were associated with substantially higher estimated energy use. The highest-scoring model, \texttt{gpt-5.5}, achieved a Best Risk score of 96.02 with estimated energy use of 6.4390 kWh per million output tokens.

Several smaller or more efficient models achieved lower but still comparatively strong safety scores at much lower estimated energy cost. For example, \texttt{claude-haiku-4.5} achieved a Best Risk score of 93.41 while consuming 0.1095 kWh per million output tokens. Compared with \texttt{gpt-5.5}, this corresponds to a 98.3\% reduction in estimated energy use for a 2.61 percentage-point reduction in Best Risk score. \texttt{gpt-4o-mini} had the lowest estimated energy use among the models shown in Table \ref{tab:models} at 0.0846 kWh per million output tokens, but also had a lower Best Risk score of 85.27. 


\textbf{Configuration-Level Variation:} The 47 configurations varied by prompt, and six models also varied by reasoning level. Across 23 matched prompt comparisons, therapeutic prompts scored higher in 11 cases and lower in 12, with differences ranging from $-2.72$ to $+6.07$ points. High reasoning scored higher than no reasoning in 4 of 12 comparisons, while low reasoning scored higher in 5 of 9. EcoLogits returned identical impact estimates within each base model because its fixed-token estimates do not distinguish configuration-specific costs. The analysis captures configuration-level variation in safety scores, but not in environmental impact.


\begin{figure*}[t]
    \centering
    \includegraphics[width=0.65\textwidth]{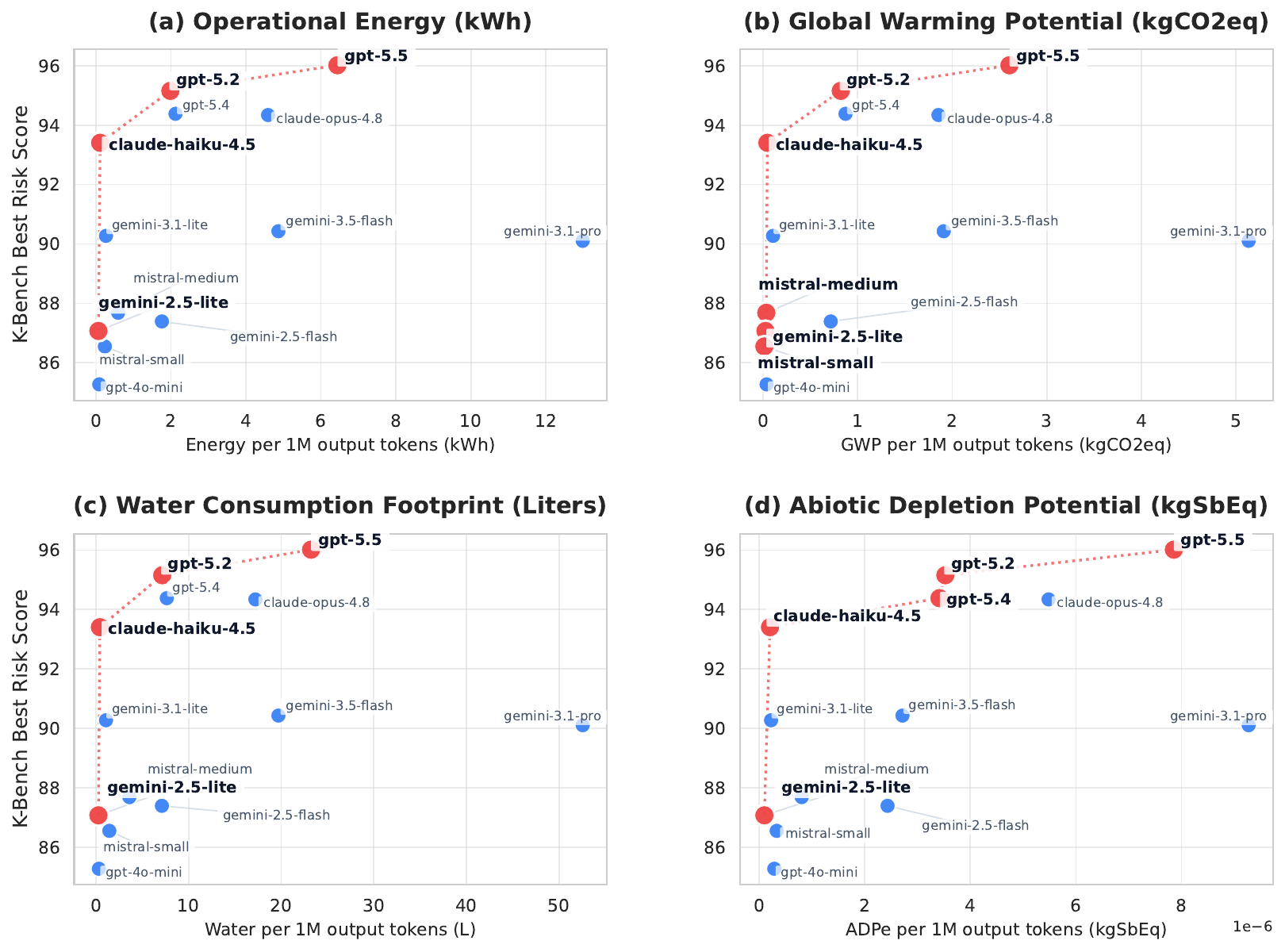}
    \caption{Clinical safety scores and estimated environmental impacts across base models. Red points identify nondominated models on the Pareto frontier for each indicator, and dotted lines connect the observed frontier points as a visual guide.}
    \label{fig:pareto_grid}
\end{figure*}

\textbf{Pareto Frontier Across Environmental Indicators:} Figure \ref{fig:pareto_grid} characterizes these trade-offs by plotting each base model's K-Bench Best Risk score against four environmental indicators: energy use, global warming potential, water consumption, and abiotic depletion potential. The scatter plots show a non-linear relationship between clinical safety scores and environmental impact. Each indicator has its own Pareto frontier, comprising models that are not outperformed by another model on both safety score and estimated environmental impact. Across the four indicators, the largest increases in estimated impact occur among models with the highest safety scores. The plots have similar shapes because EcoLogits derives these indicators from shared assumptions about model inference, hardware, and data-center operation; they should therefore not be interpreted as fully independent measures.

The same pattern was observed across the other environmental indicators. Relative to \texttt{claude-haiku-4.5}, \texttt{gpt-5.5} had an approximately 57-fold higher global warming potential, 56-fold higher water consumption, and 39-fold higher abiotic depletion estimate per million output tokens. Thus, the environmental cost of moving from a high-performing efficient model to the highest-scoring model was not limited to operational energy use, but appeared across multiple life-cycle indicators.

Taken together, these results suggest that the final increments in observed clinical safety performance are associated with disproportionately larger estimated environmental impacts. This supports treating therapeutic AI model selection as a multi-objective optimization problem rather than selecting solely by highest safety score. One strategy is dynamic model selection, using smaller, efficient models for lower-risk interactions and larger models where additional performance is clinically relevant.

\section{Discussion \& Limitations}



\textbf{Implications for Clinical AI Deployment:} The Pareto frontier can guide model selection under clinical and environmental constraints. For example, \texttt{claude-haiku-4.5} achieved a comparatively high safety score with substantially lower estimated energy use than the highest-scoring model. Its acceptability depends on use case, risk profile, monitoring, and safety margins. Aggregate K-Bench scores are comparative, not universal safety thresholds: a score of 85.27 alone does not establish safety, and small differences are not necessarily clinically meaningful. Their importance depends on whether they reflect high-risk failures, so model selection requires dimension- and case-level analysis.

\textbf{Test-Time Compute and Clinical Safety}: A secondary row-level finding concerns test-time compute and clinical safety. Additional reasoning did not consistently improve K-Bench performance. For example, \texttt{gpt-5.5} with \texttt{none reasoning} achieved the highest observed Best Risk score of 96.02, whereas the same base model with \texttt{low reasoning} achieved a lower score of 95.50. This finding should be interpreted cautiously because it compares evaluated configurations rather than all possible reasoning methods. Nevertheless, additional inference-time computation may not reliably improve clinical safety.



\textbf{Limitations:} Although efficient models performed well on the Pareto frontier, they did not achieve the highest observed safety scores in this dataset. Future research should examine whether this gap can be reduced through improved prompting, domain-specific training or fine-tuning, risk-specific monitoring, or model routing. These approaches should be evaluated for greater efficiency without compromising high-risk performance.

EcoLogits estimates rely on modeled assumptions about hardware and data-center efficiency rather than direct measurements of provider-side resource use. This limitation is especially relevant to closed systems: 15 base architectures were excluded because the required model, hardware, or routing information was unavailable. More standardized reporting of inference hardware, deployment assumptions, and life-cycle indicators would support more reliable comparisons of sustainability claims in clinical AI.

\
\section{Conclusion}

This study examined the relationship between clinical safety and estimated environmental impact in therapeutic large language models by combining K-Bench clinical safety scores with EcoLogits sustainability estimates. Across 47 supported configurations, we found that the highest observed safety scores were associated with substantially higher estimated energy use, global warming potential, water consumption, and abiotic depletion.



The analysis found that additional test-time reasoning did not consistently improve clinical safety within the evaluated configurations. This cautions against assuming that larger models or more inference-time computation will necessarily provide the most efficient route to safer therapeutic AI. Overall, the findings support treating therapeutic AI deployment as a multi-objective optimization problem, in which clinical safety, environmental impact, cost, and operational risk are evaluated together. Dynamic model selection and model cascading may offer practical ways to preserve safety in higher-risk cases while reducing the environmental footprint of routine interactions.

\begin{acks}
This study was supported by the Digital Good Network and Kivira Health.
\end{acks}

\bibliographystyle{ACM-Reference-Format}
\bibliography{software}

\end{document}